\documentclass[12pt,letter]{article}

\usepackage{graphicx, epsfig, color, cite}

\def\eslt{\not\!\!{E_T}}
\def\to{\rightarrow}

\def\bi{\begin{itemize}}
\def\ei{\end{itemize}}

\def\tst{\tilde t}

\def\tg{\tilde g}

\def\tq{\tilde q}
\def\tw{\widetilde W}
\def\tz{\widetilde Z}
\def\alt{\stackrel{<}{\sim}}
\def\agt{\stackrel{>}{\sim}}
\def\lesssim{\stackrel{<}{\sim}}
\def\gtrsim{\stackrel{>}{\sim}}
\def\be{\begin{equation}}  
\def\ee{\end{equation}}  

\newcommand\prd[3]{{\it Phys.\ Rev.\ }{\bf D #1} (#2) #3}
\newcommand\prD[3]{{\it Phys.\ Rev.\ }{\bf D #1} (#2) #3}

\newcommand\prl[3]{{\it Phys.\ Rev.\ Lett.\ }{\bf #1} (#2) #3}
\newcommand\plb[3]{{\it Phys.\ Lett.\ }{\bf B #1} (#2) #3}

\newcommand\jhep[3]{{\it J. High Energy Phys.\ }{\bf #1} (#2) #3}

\newcommand\ijmpa[3]{{\it Int.\ J.\ Mod.\ Phys.\ }{\bf A #1} (#2) #3}
\newcommand\npb[3]{{\it Nucl.\ Phys.\ }{\bf B #1} (#2) #3}

\newcommand\zpc[3]{{\it Z.\ Physik }{\bf C #1} (#2) #3}

\begin{document}

\begin{titlepage}
\vspace*{-1cm}
\flushright{UH-511-1185-12}

\vspace*{1cm}
\begin{center}
{\LARGE  
$WZ$ plus missing-$E_T$ signal\\[3mm] 
from gaugino pair production at LHC7
}\\
\vspace{1cm} \renewcommand{\thefootnote}{\fnsymbol{footnote}}
{\large Howard Baer$^{1}$\footnote[1]{Email: baer@nhn.ou.edu },
Vernon Barger$^{2}$\footnote[2]{Email: barger@pheno.wisc.edu},
Sabine Kraml$^{3}$\footnote[2]{Email: sabine.kraml@lpsc.in2p3.fr},
Andre Lessa$^{4}$\footnote[3]{Email: lessa@fma.if.usp.br},\\
Warintorn Sreethawong$^1$\footnote[4]{Email: wstan@nhn.ou.edu}
and Xerxes Tata$^5$\footnote[5]{Email: tata@phys.hawaii.edu }} \\
\vspace{0.6cm} \renewcommand{\thefootnote}{\arabic{footnote}}
{\it 
$^1$Dept. of Physics and Astronomy,
University of Oklahoma, Norman, OK 73019, USA \\
$^2$Dept.\ of Physics, University of Wisconsin, Madison, WI 53706, USA\\
$^3$Laboratoire de Physique Subatomique et de Cosmologie, UJF Grenoble 1, 
CNRS/IN2P3, INPG, 53 Avenue des Martyrs, F-38026 Grenoble, France\\
$^4$Instituto de F\'isica, 
Universidade de S\~ao Paulo, S\~ao Paulo-SP, Brazil\\
$^5$Dept.\ of Physics and Astronomy, University of Hawaii, Honolulu, HI 96822, USA\\
}

\end{center}

\vspace{0.3cm}
\begin{abstract}
\noindent 
LHC searches for supersymmetry currently focus on strongly produced
sparticles, which are copiously produced if gluinos and squarks have
masses of a few hundred GeV. However,
in supersymmetric models with heavy scalars, as favored by the
decoupling solution to the SUSY flavor and $CP$ problems, and
$m_{\tg}\agt 500$ GeV as indicated by recent LHC results,
chargino--neutralino $(\tw_1^\pm\tz_2$) production is the dominant cross
section for $m_{\tw_1} \sim m_{\tz_2} < m_{\tg}/3$ at LHC with
$\sqrt{s}=7$ TeV (LHC7).  Furthermore, if 
$m_{\tz_1}+m_Z \lesssim m_{\tz_2}\lesssim m_{\tz_1}+m_h$, then
$\tz_2$ dominantly decays via  $\tz_2\to\tz_1 Z$, while $\tw_1$ decays via $\tw_1\to \tz_1 W$. 
We investigate the LHC7 reach in the $WZ +\!\eslt$ channel
(for both leptonic and hadronic decays of the $W$ boson) 
in models with and without the assumption of gaugino mass universality. 
In the case of the mSUGRA/CMSSM model with heavy squark masses, 
the LHC7 discovery reach in the $WZ+\!\eslt$ channel becomes competetive with 
the reach in the canonical $\eslt$ + jets channel for integrated luminosities $\sim 30$~fb$^{-1}$.
We also present the LHC7 reach for a simplified model with arbitrary $m_{\tz_1}$
and $m_{\tw_1} \sim m_{\tz_2}$. Here, we find a reach of up to $m_{\tw_1}\sim 200~(250)$~GeV 
for 10~(30)~fb$^{-1}$.
\vspace{0.8cm}


\end{abstract}


\end{titlepage}

\section{Introduction}
\label{sec:intro}

A major goal of the CERN Large Hadron Collider (LHC) is to test the idea
of weak scale supersymmetry (SUSY)\cite{wss}, wherein superpartners of
the Standard Model (SM) particles have masses of the order of 1~TeV.
The SUSY searches by the ATLAS and CMS collaborations have reported no
signal beyond SM expectations~\cite{atlas,cms} in $\sim 1$~fb$^{-1}$ of
data.
Interpreting their results within the mSUGRA/CMSSM model~\cite{msugra}, 
ATLAS and CMS exclude roughly the
mass range $m_{\tq}\sim m_{\tg}\alt 1$~TeV for $m_{\tq}\simeq m_{\tg}$, and 
$m_{\tg}\alt 550$~GeV in the case where $m_{\tq}\gg m_{\tg}$.\footnote{To be precise, 
in the mSUGRA/CMSSM interpretation, squark masses are varied up to  
$m_{\tq}\alt 2$~TeV, giving a gluino mass limit of about $m_{\tg}\agt 700$~GeV; 
this limit suffers further weakening for decoupling scalars: see \cite{atlasSO10}.}
This reach will soon 
be extended since each experiment now has $\sim 5$ fb$^{-1}$ of data collected. Analysis of this
extended data sample is eagerly anticipated by the HEP community. 

Within a large class of SUSY models, it is expected that pair production
of strongly interacting
sparticles---$\tg\tg$, $\tg\tq$ and $\tq\tq$ production---constitutes  
the dominant SUSY production cross sections\cite{bcpt1,bcpt2}. 
The gluinos and squarks are then expected to decay through 
a (possibly lengthy) cascade to lighter sparticles plus
SM particles, until the decay chain terminates in 
the (stable) lightest SUSY particle (LSP) \cite{cascade}.  The LSP is expected
from cosmological arguments to be a massive, neutral, weakly interacting
particle (such as the lightest neutralino $\tz_1$) and so does not deposit
energy in the experimental apparatus, giving rise to the classic
missing transverse energy ($\eslt$) signature. 
Thus, gluino and 
squark pair production followed by cascade decays is expected to give rise to final states
containing multiple isolated leptons, multiple jets and
$\eslt$~\cite{btw}. 

While weak scale supersymmetric models are theoretically very
compelling, they do suffer from a variety of problems, including 1.\ the
SUSY flavor problem, 2.\ the SUSY $CP$ problem, 3.\ the gravitino
problem, and 4.\ the danger of too rapid proton decay in SUSY grand
unified theories (GUTs). All four of these problems are greatly
ameliorated if not solved by the decoupling solution, wherein first and
second generation sfermion masses are pushed into the multi-TeV
regime or even beyond. Naturalness may be maintained in models wherein
sparticles that couple directly to the Higgs sector---the third
generation scalars and electroweak-inos---remain at or below the TeV
scale\cite{dine,ckn}.  Also, in many SUSY models, it is expected that
gaugino mass parameters unify at the GUT scale, in parallel with
unification of gauge couplings. Renormalization group running effects
result in weak scale gaugino masses occurring in the approximate ratio
$M_1:M_2:M_3\sim 1:2:7$. We would thus expect the physical gluino
$\tg$, the wino-like chargino $\tw_1$ and the bino-like neutralino
$\tz_1$ to be found with roughly the same mass ratio, provided the
superpotential $\mu$-parameter $|\mu | \gg M_{2}$. Consequently, in
models with gaugino mass unification, the experimental bounds on the
gluino mass impose severe constraints on chargino and neutralino masses.
Current analyses do not put independent constraints on the
electroweak-ino masses if the gaugino mass unification condition is
dropped \cite{Sekmen:2011cz}.  Moreover, the relative strengths of
signals in various multilepton topologies (as well as the gluino mass
reach if the parent-daughter mass difference is sufficiently small)
depend sensitively on the $\tg-\tz_1$ and/or $\tg-\tw_1$ mass
differences. Finally, an independent discovery of directly produced
charginos and neutralinos is essential to elucidate the supersymmetry
origin of any excess in the well-studied multilepton plus multijet plus
$\eslt$ channel at the LHC.  It is therefore interesting and relevant to
find ways to discover charginos and neutralinos independently of
gluinos.

Another point is important to note: as we push the gluino mass to larger
values, convolution of the $\tg\tg$ subprocess cross sections with
parton distribution functions (PDFs) requires sampling higher and higher
values of parton fractional momentum $x_F$. For such high values of
$x_F$, the parton-parton luminosity is sharply falling.  At some point
we expect that, despite being strongly-produced, gluino pair production
will no longer dominate over electroweak-ino pair production, since
these latter reactions will sample the PDFs at much lower values of
$x_F$ if electroweak-inos are significantly lighter than gluinos.

\begin{figure}[t]
\begin{center}
\epsfig{file=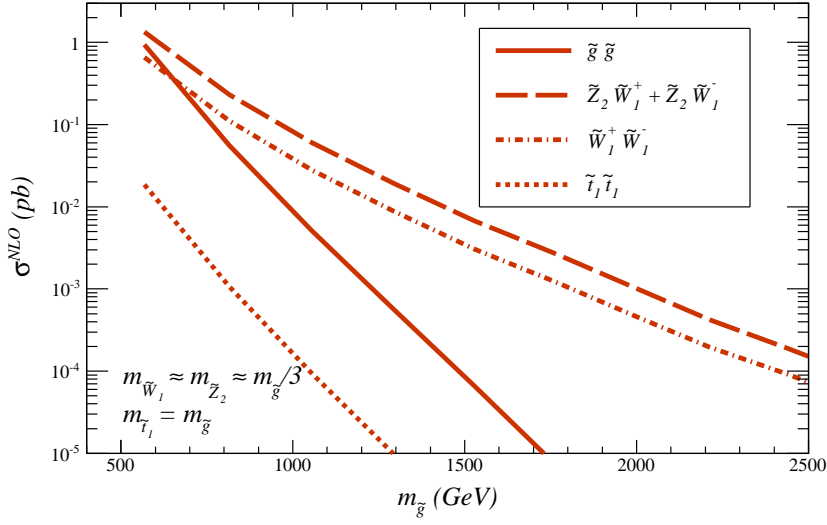,width=12cm}
\end{center}
\vspace*{-4mm}
\caption{\it Total NLO cross sections (from Prospino) for $\tg\tg$,
$\tw_1^\pm\tz_2$ and $\tw_1^+\tw_1^-$ production at LHC7 versus
$m_{\tg}$, where $m_{\tq}=15$ TeV and $m_{\tw_1} \approx m_{\tz_2}
\approx m_{\tg}/3$. The dotted line shows the cross section for $\tst_1
\bar{\tst_1}$ production with $m_{\tst_1}=m_{\tg}$ and neglecting intra-generational
squark mixing.
\label{fig:xsecs}}
\end{figure}

To illustrate this, we plot in Fig.~\ref{fig:xsecs} the $\tg\tg$,
$\tw_1^\pm\tz_2$ and $\tw_1^+\tw_1^-$ production cross sections in 
pb at LHC with $pp$ collisions at $\sqrt{s}=7$ TeV. Our results are in
NLO QCD from the program Prospino\cite{prospino}. We take $m_{\tq}\simeq
15$ TeV for the first and second generations, in accord with a
decoupling solution to the above-mentioned pathologies and, for
simplicity, assume universal gaugino masses at the GUT scale.  From
Fig.~\ref{fig:xsecs}, we see that gluino-pair production is dominant for
$m_{\tg}\alt 500$ GeV.  For higher values of $m_{\tg}$, $\tw_1^\pm\tz_2$
production is dominant, followed by $\tw_1^+\tw_1^-$ production (the
reaction $\tw_1^\pm\tz_1$ has lower cross section,\footnote{For the
wino-like $\tw_1$ and $\tz_2$, $\tw_1\tz_2$ production occurs via the
unsuppressed isotriplet $W\tw_1\tz_2$ gauge coupling, whereas the
$W\tw_1\tz_1$ coupling is strongly suppressed because it arises only due
to the subdominant higgsino content of the wino-like chargino and the
bino-like neutralino --- the $W$-bino-wino coupling is forbidden by gauge
invariance.} as can be seen {\it e.g.}\  in Fig.~12.23 of
Ref.~\cite{wss}).  For LHC with $\sqrt{s}=14$ TeV, $\tg\tg$ production remains
dominant up to $m_{\tg}\sim 1$ TeV if squarks are very heavy.  Since
ATLAS and CMS already exclude $m_{\tg}\alt 550$ GeV when $m_{\tq}$ is
large it may prove fruitful to probe electroweak gaugino pair production
in the 2011 data but most of all in the 2012 LHC run.  This was recognized 
early on in \cite{bcpt1,bcpt2} and also more recently in 
in~\cite{Mrenna:2011ek,wh}.  Recognizing that the
stability of the Higgs sector also requires sub-TeV top squarks, we also
show the cross section for top squark pair production for
$m_{\tst_1}=m_{\tg}$ by the dotted line\footnote{The LO top squark pair
production cross section is determined by QCD and is independent of
$m_{\tg}$. In other words, for the dotted line, the graph is plotted
versus $m_{\tst_1}$. If other third generation squarks are also light,
their pair production cross sections are also given by the dotted line
with the understanding that the label on the horizontal axis is the
corresponding squark mass.} in Fig.~\ref{fig:xsecs}. We see that this
cross section also drops off rapidly with the top squark mass. Unless
top squarks are exceptionally light (with masses of order $m_{\tw_1}$ or
smaller, and certainly much smaller than $m_{\tg}$), electroweak-ino
production remains the dominant mechanism.

Let us next examine the signatures resulting from $\tw_1\tz_2$
production. If $m_{\tz_2}< M_Z + m_{\tz_1}$, the well-known trilepton
signal provides a golden signature for chargino-neutralino
production\cite{trilepton,bcpt2} provided only that the branching fraction for
neutralino decay is not unduly suppressed\cite{supp}.
The two-body chargino decay $\tw_1\to\tz_1 W$ is expected to dominate  for
$m_{\tw_1}\agt M_W +m_{\tz_1}$, while the two-body decay
$\tz_2\to\tz_1 Z$ dominates for $M_Z+m_{\tz_1} \lesssim m_{\tz_2}\lesssim m_h +m_{\tz_1}$.  
For even higher values of
$m_{\tz_2}$, i.e.\ $m_{\tz_2}\agt m_{\tz_1} + m_h $, the decay mode $\tz_2\to\tz_1 h$  turns on and dominates.

\begin{figure}[!t]
\begin{center}
\epsfig{file=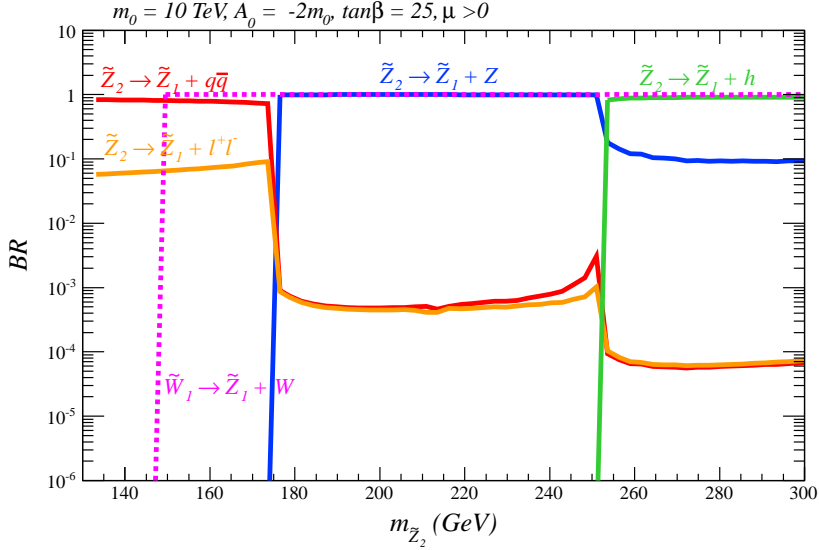,width=12cm}
\end{center}
\vspace*{-4mm}
\caption{\it Some prominent branching fractions for $\tz_2$ decay in the mSUGRA model
  with parameters $m_0=10$ TeV, $A_0=-2m_0$, $\tan\beta =25$ and $\mu >0$. We also 
  show the $\tw_{1} \to W + \tz_1$ branching fraction (dotted line).
\label{fig:BFs}}
\end{figure}

This is illustrated in Fig.~\ref{fig:BFs}, where we show the $\tz_2$ branching fractions versus
$m_{\tz_2}$ for a mSUGRA model line with $m_0=10$ TeV, $A_0=-2m_0$,
$\tan\beta =25$ and $\mu >0$.  We vary $m_{1/2}$ to obtain the
variation in $m_{\tz_2}$.  In this case, $\tw_1 \tz_2 \to WZ + \tz_1\tz_1$ is 
kinematically allowed for 175~GeV $\lesssim m_{\tz_2} \lesssim
250$~GeV, which corresponds to gluino masses in the interval
600~GeV~$\lesssim m_{\tg} \lesssim 800$~GeV. Thus, in this mass range,
we expect the single reaction $pp\to\tw_1\tz_2$ followed by
$\tw_1\to\tz_1 W$ and $\tz_2\to\tz_1 Z$ to be the dominant SUSY production
and decay process at LHC7 for models with full gaugino mass unification.  The
endpoints of this interval can shift up or down in
non-universal mass scenarios.

\section{Trilepton$+\eslt$ channel}
\label{sec:3l}

We begin by examining the viability of the reaction 
$pp\to\tw_1\tz_2\to WZ+\!\eslt$ for SUSY discovery at LHC7,  
focusing on the case where both $Z$ and $W$ decay leptonically, 
resulting in clean trilepton events.
 It is worth mentioning that the trilepton signal from the decay
$\tz_2\to \tz_1Z$ where a pair of opposite-sign same-flavor (OS/SF) dileptons
reconstruct the $Z$ mass has generally been regarded as unobservable
because of large SM background from $WZ$ production. The case where the
$W$ decays hadronically will be discussed in Section~\ref{sec:lljj}.

For our LHC7 event generation, we use the event generator Isajet
7.79\cite{isajet} for signal reactions, while for the simulation of the
background events, we use AlpGen\cite{alpgen} and MadGraph\cite{madgraph} to compute the hard
scattering events and Pythia \cite{pythia} for the subsequent showering
and hadronization.  In our simulation, we include the following
backgrounds for the $WZ+\eslt$ signal: $t\bar{t}$, $W(\ell \nu)W(\ell \nu)$, $W(\ell \nu)Z(\ell \ell)$, $ZZ$,
$W(\ell \nu)+tb$, $Z(\ell \ell)+jets$, $W(\ell \nu)+jets$, $Z(\ell \ell)+b\bar{b}$,
$Z(\ell \ell)+t\bar{t}$ and $W+t\bar{t}$. For $t\bar{t}$,
$Z+jets$, $W+jets$, $Z + b\bar{b}$ and $Z + t\bar{t}$ we include the
full matrix elements for at least two real parton emissions and use the
MLM matching algorithm to avoid double counting. For $WZ$ production we include the full matrix elements
for the $2 \to 4$ process $pp \to WZ \to \ell^+ \ell^- \ell^{'} \nu^{'}$. K-factors for both
signal and background\footnote{For the background processes where the NLO cross section is not known
we take the K-factor to be 1.} (BG) are included and are computed using
Prospino\cite{prospino} and MCFM\cite{mcfm}, respectively.

In our calculations, we employ a toy detector simulation with
calorimeter cell size $\Delta\eta\times\Delta\phi=0.05\times 0.05$ and
$-5<\eta<5$ . The HCAL (hadronic calorimetry) energy resolution is taken
to be $80\%/\sqrt{E}\oplus 3\%$ for $|\eta|<2.6$ and FCAL (forward
calorimetry) is $100\%/\sqrt{E} \oplus 5\%$ for $|\eta|>2.6$, where the two
terms are combined in quadrature. The ECAL (electromagnetic calorimetry)
energy resolution is assumed to be $3\%/\sqrt{E}\oplus 0.5\%$. In all
these, $E$ is the energy in GeV units. We use the
cone-type Isajet \cite{isajet} jet-finding algorithm to group the
hadronic final states into jets. Jets and isolated lepton are defined as
follows: 
\bi 
\item Jets are hadronic clusters with $|\eta| < 3.0$,
$R\equiv\sqrt{\Delta\eta^2+\Delta\phi^2}\leq0.4$ and $E_T(jet)>40$ GeV.
\item Electrons and muons are considered isolated if they have $|\eta| <                                   
2.5$, $p_T(l)>10 $ GeV with visible activity within a cone of $\Delta                                      
R<0.2$ about the lepton direction, $\Sigma E_T^{cells} < \min[5,0.15p_{T}(l)]$ GeV.
\item  We identify hadronic clusters as
$b$-jets if they contain a $B$ hadron with $E_T(B)>$ 15 GeV, $|\eta(B)|<$ 3.0 and
$\Delta R(B,jet)<$ 0.5. We assume a tagging efficiency of 60$\%$ and
light quark and gluon jets can be mis-tagged
as a $b$-jet with a probability 1/150 for $E_{T} \leq$ 100 GeV,
1/50 for $E_{T} \geq$ 250 GeV, with a linear interpolation
for 100 GeV $\leq E_{T} \leq$ 250 GeV.
\ei

Next, we invoke the following pre-selection cuts on our 
signal and background event samples to extract those with a 
$\ell^+\ell^- \ell^{'} + \eslt$ topology: 
\begin{itemize}
\item[] \underline{Pre-Selection Cuts:}
\item $n(b-jets)=0$ (to aid in vetoing $t\bar{t}$ background),
\item 3 isolated leptons with $p_T(\ell )>20$ GeV and
\item $|m(\ell^{+}\ell^{-}) - M_Z| < 10$~GeV,
\end{itemize}
where two of the leptons in the event must form an OS/SF pair. If more than one OS/SF pairing
is possible, the pair which minimizes $|m(\ell^{+}\ell^{-}) - M_Z|$ is chosen. The remaining lepton is labeled $\ell^{'}$.

\begin{figure}[!t]
\begin{center}
\epsfig{file=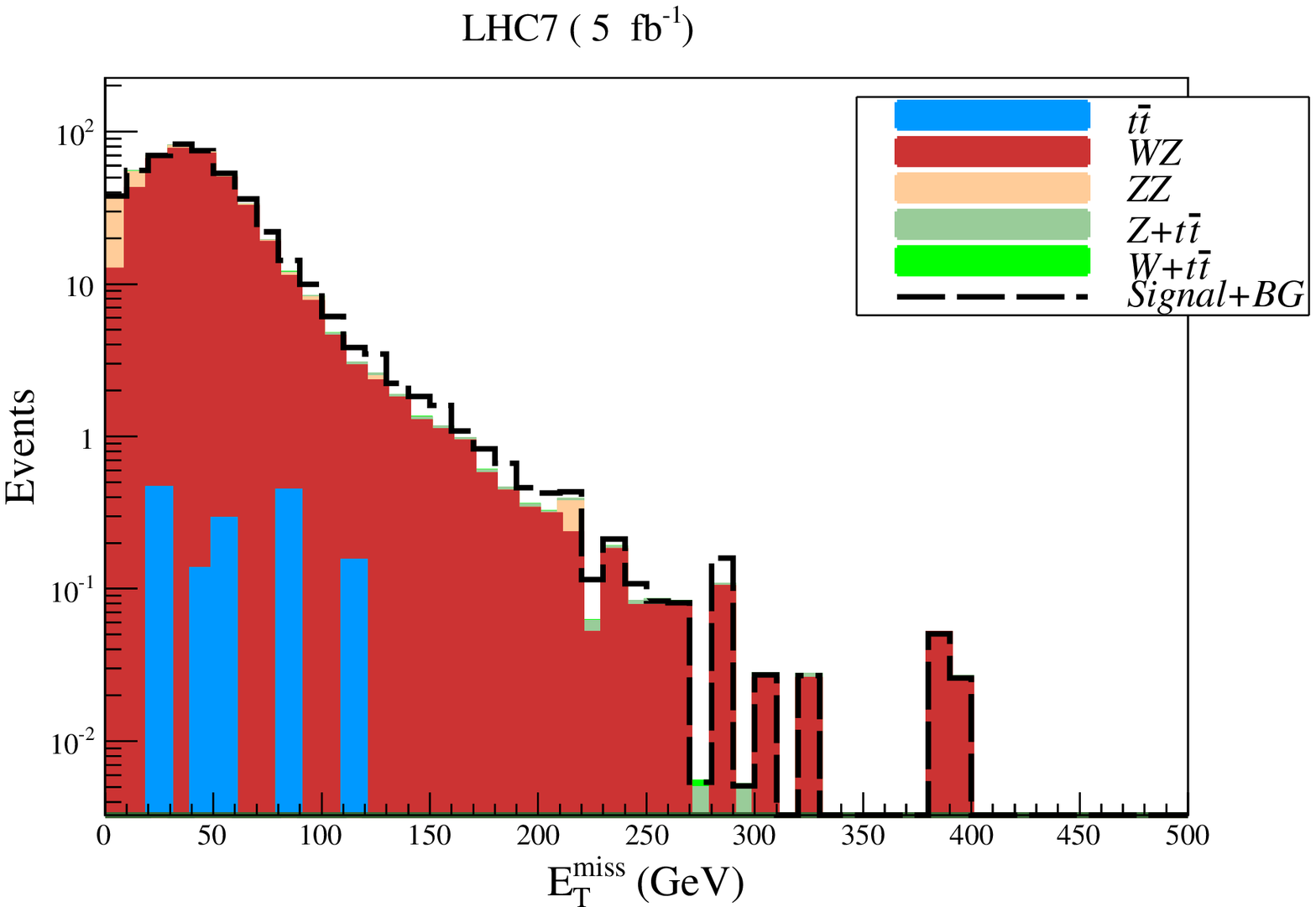,width=10cm}
\epsfig{file=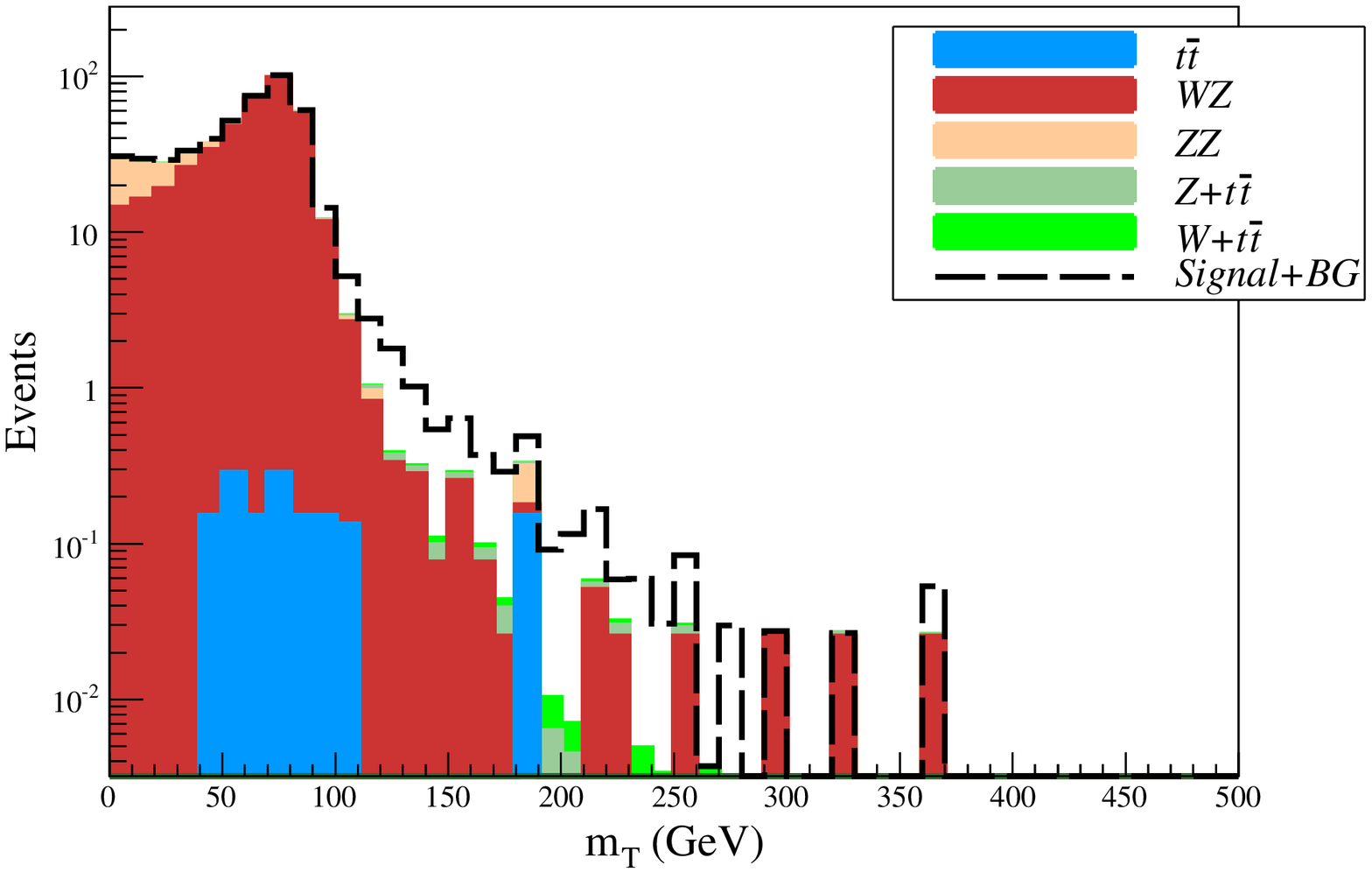,width=10cm}
\end{center}
\vspace*{-4mm}
\caption{\it $\eslt$ and transverse mass ($m_T(\ell^{'}),\eslt$) distributions
in $3\ell\,+\!\eslt$ events after pre-selection cuts, 
for an integrated luminosity of 5~fb$^{-1}$.  The summed SM
backgrounds are shaded while the signal plus background is shown by the
dashed histogram. Only the
dominant background processes are shown.  The signal point has $m_{\tw_1} =
189.3$~GeV, $m_{\tz_2} = 187.3$~GeV and $m_{\tz_1} = 89.4$~GeV.
}\label{fig:MET}
\end{figure}

In Fig.~\ref{fig:MET} we show the $\eslt$ and transverse mass
($m_{T}(\ell^{'},\eslt )$) distributions
for the signal and the SM BG after the pre-selection cuts have been
applied.  The signal point has $m_{\tw_1} = 189.3$~GeV, $m_{\tz_2} =
187.3$~GeV and $m_{\tz_1} = 89.4$~GeV and we only consider $\tw_1\tz_2$ production.  Due to its
relatively light parent mass scale, the signal presents a soft $\eslt$
spectrum, barely visible above the SM background. This is in strong
contrast with events from production of the much heavier gluinos or squarks, where the cascade
decays to the LSP result in a usually much harder $\eslt$ spectrum.
Therefore, the usual $\eslt$ plus jets/leptons searches (optimized to
look for strongly produced gluinos and squarks) are insensitive to the
$\tw_1 \tz_2$ signal.

As seen in the upper frame of Fig.~\ref{fig:MET}, after the
pre-selection cuts the BG is dominated by $ZZ$ production at low
$\eslt$ and by $WZ$ production for $\eslt \gtrsim 20$~GeV.  The
transverse mass $m_T(\ell',\eslt )$ from $W\to\ell'\nu_{\ell'}$, shown
in the lower frame of Fig.~\ref{fig:MET}, falls sharply beyond the
expected Jacobian peak at $m_T = M_W$. In constrast, the corresponding
signal distribution from $\tw_1\tz_2$ production extends to
considerably larger values due to the presence of the two neutralinos
in the final state. Therefore, a $m_{T}$ cut is extremely efficient to
suppress the $WZ$ background.  This is seen in the lower frame of
Fig.~\ref{fig:MET}, where the signal distribution clearly stands out for $m_T > 100$~GeV.
However, since a precise prediction for the $m_T$ tail
from $WZ$ events requires a full detector simulation or data-driven
estimates, we define a conservative signal region requiring:
\begin{itemize}
\item $\eslt >50$ GeV,
\item $m_{T}(\ell^{'},\eslt ) > 125$ GeV .
\end{itemize}

The BG cross sections from the dominant SM processes after each of the
cuts mentioned above, together with the corresponding cross sections for
the representative signal point with $m_{\tw_1} = 189.3$~GeV, $m_{\tz_2}
= 187.3$~GeV and $m_{\tz_1} = 89.4$~GeV, are shown in Table
\ref{tab:xsecs}. We stress that the signal shown in Fig.~\ref{fig:MET}
and listed in Table~\ref{tab:xsecs} comes exclusively from
$\tw_1\tz_2$ production. Depending on the sparticle spectrum, the
actual signal may be larger if heavier electroweak-inos are also
accessible, or if gluino and/or squark pair production followed by their
cascade decays to the $WZ$ final state is sizeable.  Nonetheless, a
trilepton signal would be visible with an integrated luminosity of $\sim
10$~fb$^{-1}$ at LHC7
even if light electroweak-inos are the {\it only} SUSY particles being
produced.

\begin{table}
\begin{center}
\begin{tabular}{|l|r|r|r|r|r|r|r|}
\hline
& $t\bar{t}$ & $WZ$ & $ZZ$ & $Z+t\bar{t}$ &  $W+t\bar{t}$ & Total BG & Signal\\
\hline
Events Generated & 5.1M & 100K & 194K & 451K & 9.5M & & 200K \\
Total $\sigma$ (fb) & $1.6 \times 10^5$ & $5.1 \times 10^2$ & $5.4 \times 10^3$ & 22.3 & 183 & $7.8 \times 10^{6}$ &  $1.1 \times 10^4$ \\
\hline
$n(b) = 0, n(l) = 3$  & 1.6 & 85.1 & 9.2 & 0.9 & 0.4 & 97.5 & 6.7 \\
OS/SF pair & 1.1 & 84.9 & 9.2 & 0.9 & 0.3 & 96.6 & 6.7 \\
$m(\ell^{+}\ell^{-})$ cut & 0.3 & 79.1 & 9.1 & 0.66 & 0.06 & 89.5 & 6.6 \\
$m_{T} > 125$~GeV & 0.03 & 0.20 & 0.03 & 0.03 & 0.02 & 0.31 & 0.67 \\
$\eslt > 50$~GeV & 0.03 & 0.17 & 0 & 0.03 & 0.02 & 0.25 & 0.64 \\
\hline
\end{tabular}
\caption{Number of events generated, total cross section and
cross section after cuts for the dominant backgrounds in the trilepton
channel and for the signal.  All cross sections are in fb and the
signal is from just $\tw_1\tz_2$ production with $m_{\tw_1} =
189.3$~GeV, $m_{\tz_2} = 187.3$~GeV and $m_{\tz_1} = 89.4$~GeV.  The
Total BG values include all processes listed in the text, including the
subdominant ones not shown in the Table.
\label{tab:xsecs}}
\end{center}
\end{table}
%

\subsection{LHC7 Reach}

\begin{figure}[t]
\begin{center}
\epsfig{file=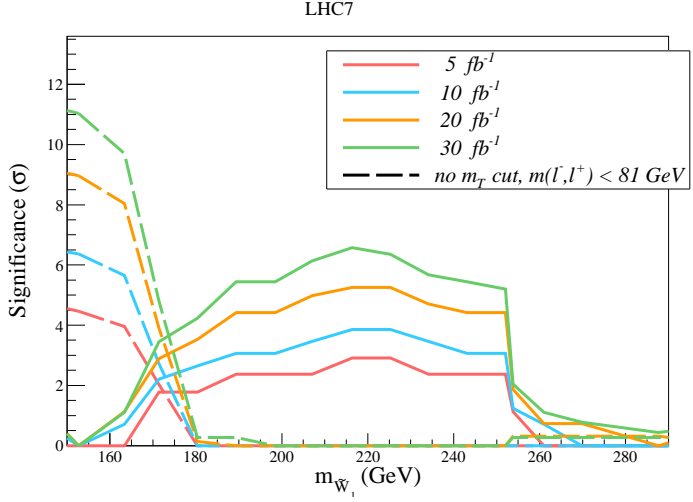,width=10cm}
\end{center}
\vspace*{-4mm}
\caption{\it 
Significance of $\tw_1\tz_2\to WZ+\!\eslt\to 3\ell +\!\eslt$ signal for
various integrated luminosities at LHC7.
The solid lines have all the trilepton cuts listed in Table~\ref{tab:xsecs}, while the dashed
lines do not include the $m_{T}$ cut and require $M_Z - m(\ell^{+}\ell^{-}) > 10$~GeV instead.
}\label{fig:sig-3l} 
\end{figure}

As shown in Table~\ref{tab:xsecs} and Fig.~\ref{fig:MET}, for $m_{\tw_1}
= 189.3$~GeV, $m_{\tz_2} = 187.3$~GeV and $m_{\tz_1} = 89.4$~GeV, only
an excess of $\sim 2$ events in the trilepton channel (after cuts) would
be expected for luminosity of $\sim 5$~fb$^{-1}$.  Thus larger integrated luminosities
are required in order to claim a signal.
In Fig.~\ref{fig:sig-3l}, we show the signal significance for various
integrated luminosities versus $m_{\tw_1}$ (solid lines).
For now we use a mSUGRA model line with $m_0 = 10$~TeV, $A_0 = -2m_0$,
$\tan\beta = 25$ and $\mu >0$, and we consider the signal only from
$\tw_1\tz_2$ production.
To allow for the low signal rates, the significance is computed using
 Poisson statistics.  
For $m_{\tw_1} \lesssim 170$~GeV,
the decay into real $Z$s is kinematically forbidden-- as shown in
Fig.~\ref{fig:BFs}-- and the signal significance (solid lines) sharply drops in this
region.  In this case, however, the well-studied trilepton signal
mentioned earlier from $\tw_1\tz_2\to 3\ell +\eslt$ where
$m(\ell^+\ell^-)< M_Z$ is observable. To illustrate this, 
we show by dashed lines the signal significance, where the same cuts
listed in Table~\ref{tab:xsecs} are applied, except for the $m_{T}$ and $m(\ell^{+}\ell^{-})$ cuts.
Since in this region $\tz_2$ and $\tw_1$ can decay to off-shell $Z$s and $W$s we require instead:
\bi
\item $m_{T} > 0$,\ \ \  $m(\ell^{+}\ell^{-}) < M_Z -  10$~GeV. 
\ei 
As seen from Fig.~\ref{fig:sig-3l}, we confirm that the signal in the low
$m_{\tw_1}$ region ($\lesssim 170$~GeV) is readily observable via this
``golden" trilepton channel, due to the large $\tw_1\tz_2$ production
cross sections and small background.\footnote{The valley at the intersection of the solid and dashed
lines in Fig.~\ref{fig:sig-3l} arises because we have different analysis
cuts for the two-body and three-body decays of $\tz_2$. This valley
would be smoothed out (and partially filled in) in a treatment that
treats $Z$ as a resonance rather than a particle with a definite mass.}

As $m_{\tw_1}\simeq m_{\tz_2}$ increases so that the $\tz_2 \to \tz_1 Z$ decay
turns on, the significance for our $WZ\to 3\ell\,+\!\eslt$ signal increases,
reaching its maximum for $m_{\tw_1} \sim 220$~GeV.  This is due to the
fact that, for $m_{\tw_1} \lesssim 200$~GeV, $m_{\tz_2}-m_{\tz_1} - M_Z
\lesssim 15$~GeV and the $\tz_1$'s coming from $\tz_2$ decays (and to 
some extent also those from $\tw_1$ decay) are rather soft and so contribute
relatively little to both $\eslt$ and to $m_T$.  As a result, the $\eslt
> 50$~GeV and $m_{T} > 125$~GeV requirements significantly reduce the
signal in this region. As $m_{\tw_1}$ increases beyond 220~GeV, the $\tw_1
\tz_2$ production cross section (after cuts) decreases, and so does the signal
significance. Finally, once $m_{\tz_2} > m_{\tz_1} + m_{h}$ (at
$m_{\tw_1} \sim 255$~GeV), the $\tz_2 \to \tz_1 h$ decay turns on and
dominates\footnote{This decay occurs via the
Higgs-higgsino-gaugino coupling and so is suppressed by the higgsino
content of just one of the two neutralinos. In contrast, the decay to $Z$
occurs via the doubly suppressed higgsino content of {\it both}
neutralinos.} causing 
the signal to drop sharply.  

We remark that for 5~fb$^{-1}$ of data, we would expect a 2$\sigma$
effect over essentially the entire region where the decay $\tz_2\to \tz_1Z$
dominates.  Therefore, the LHC experiments already have accumulated
enough luminosity to probe this entire region at $\sim 95\%$ C.L.!
However, in the happy circumstance that some excess
is seen in the data, $\sim 20-30$~fb$^{-1}$ of data will be required in
order to establish a $5\sigma$ discovery. This may indeed be achieved 
in the 2012 run of LHC7. We note further that the SUSY
signal events will contain a distinctive asymmetry of trilepton charges
$+(+-)\ vs.\ -(+-)$ (where the $(+-)$ pair reconstructs $m_Z$) that
originates from the PDFs since LHC is a $pp$ collider. In contrast, SM
backgrounds from $t\bar{t}$ and $Zt\bar{t}$ (but not $WZ$) should have the
number of $+(+-)$ events equal to $-(+-)$ events, up to statistical fluctuations.
In addition, should a large enough data sample be accrued, the $p_T(Z)$ distribution
should be well-suited for a $\tz_2$ mass extraction since the production and decay modes
are single channel.

\begin{figure}[t]
\begin{center}
\epsfig{file=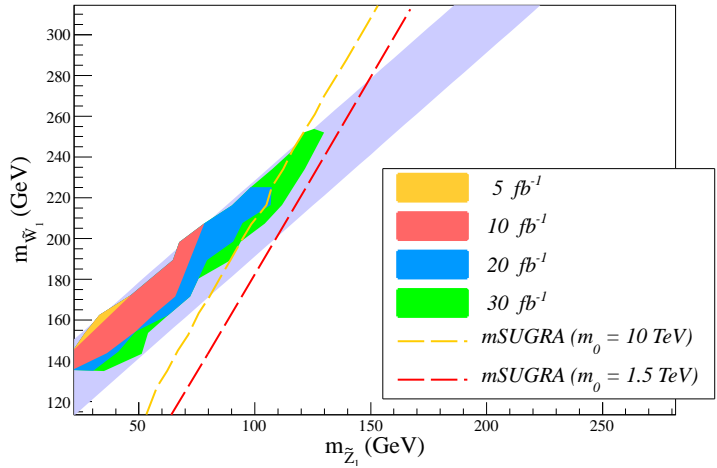,width=10cm}
\end{center}
\vspace*{-4mm}
\caption{\it $5\sigma$ discovery regions for various integrated
luminosities at LHC7 in the $m_{\tw_1}-m_{\tz_1}$ plane. We assume
$m_{\tz_2}=m_{\tw_1}$ and consider only $\tw_1\tz_2$ production. The
Higgs boson  mass is assumed to be $128.5$~GeV throughout the plane. The orange
(red) line shows the mSUGRA line with $m_0 = 10$~TeV, $A_0 = -2m_0$,
$tan\beta = 25$ and $\mu >0$ ($m_0 = 1.5$~TeV, $A_0 = 0$, $\tan\beta =
45$ and $\mu > 0$).  }\label{fig:plane-3l}
\end{figure}

In Fig.~\ref{fig:plane-3l}, we generalize our results to models with
unrelated $\tw_1$ and $\tz_1$ masses, {\it i.e.}\  models without gaugino
mass universality, taking $m_{\tz_2}=m_{\tw_1}$ and $\mu\gg M_2$.  In this figure, we
show the discovery regions for several integrated luminosities.  We
require the following discovery criteria: \bi
\item significance $> 5\sigma$, 
\item signal/BG$>0.2$ and 
\item at least 5 signal events.  
\ei 
The mSUGRA model line with $m_0 = 10$~TeV, $A_0 = -2m_0$, $\tan\beta = 25$ and $\mu>0$,
assumed in Fig.~\ref{fig:sig-3l}, is shown as the dashed orange
line.  The purple band shows the kinematically allowed region, where
$M_Z < m_{\tz_2} - m_{\tz_1} < m_h$.  As can be seen, chargino masses up
to $\sim 170$~GeV can already be probed with 5~fb$^{-1}$, if $m_{\tz_1}
\lesssim 50$~GeV.  As discussed above, for 
heavier ${\tz_1}$, the $m_{\tz_2}-m_{\tz_1}$ mass gap reduces, resulting
in softer $m_T$ and $\eslt$ distributions.  Therefore the signal
efficiency is reduced, requiring higher luminosities in order to achieve 
$5\sigma$ significance.  This effect is seen throughout the $m_{\tw_1}$ vs.\ $m_{\tz_1}$ plane, 
rendering the narrow region close to $m_{\tz_2} - m_{\tz_1} \sim M_Z$, where the $\tz_1$ is
produced at low $p_T$, inaccessible even for $\mathcal{L} =
30$~fb$^{-1}$. On the other hand, the region where $m_{\tz_2} -
m_{\tz_1} \lesssim m_h$ results in boosted $\tz_1$s and can be easily
probed until the decay $\tz_2 \to \tz_1 + h$ turns on, 
c.f.\ Fig.~\ref{fig:sig-3l}.
The 30~fb$^{-1}$ reach extends up to $m_{\tw_1} \sim 250$~GeV, for
$m_{\tz_1} \lesssim 130$~GeV, covering almost all of the kinematically
allowed region for the mSUGRA line with $m_0 = 10$~TeV, $A_0 = -2m_0$.
We also show in Fig.~\ref{fig:plane-3l} a second mSUGRA line with $m_0 =
1.5$~TeV, $A_0 = 0$, $\tan\beta = 45$ and $\mu > 0$. For these choice of
parameters the $m_{\tz_2}-m_{\tz_1}$ mass difference is reduced, due to
a small (positive) $A_0$ value and smaller squark masses.  As a result, all of the region where
the $WZ+\eslt$ channel is open falls into the inaccessible region at
high $m_{\tz_1}$.  However, values of $A_0\sim 0$ now seem excluded in mSUGRA
if indeed $m_h$ turns out to be $\sim 125$ GeV\cite{bbm}.

Up to now we have only considered 
$\tw_1\tz_2$ production. Despite having subdominant production
cross sections, production of heavier chargino $\tw_2$ and neutralinos
$\tz_{3,4}$ usually leads to a harder $\eslt$ spectrum due to their
cascade decay, possibly enhancing the signal. Furthermore, for low
$m_{0}$ ($m_{1/2}$) squark (gluino) production and cascade decay can
also enhance the trilepton signal. In order to clearly see these effects 
we choose the $A_0$ and $\tan\beta$ values from the red curve in Fig.~\ref{fig:plane-3l} ($A_0 = 0$ and $\tan\beta = 45$),
where we do not expect the $\tw_1\tz_2$ signal to be visible for any value of $m_{1/2}$, even for 30 fb$^{-1}$.
However, now we perform a scan over the $m_0-m_{1/2}$ plane
and include the production from
{\it all} SUSY particles, including squarks and gluinos.
For each point in parameter space, we
apply the trilepton cuts shown in Table~\ref{tab:xsecs} and take the
point to be visible if the discovery criteria listed above are
satisfied.

\begin{figure}[t]
\begin{center}
\epsfig{file=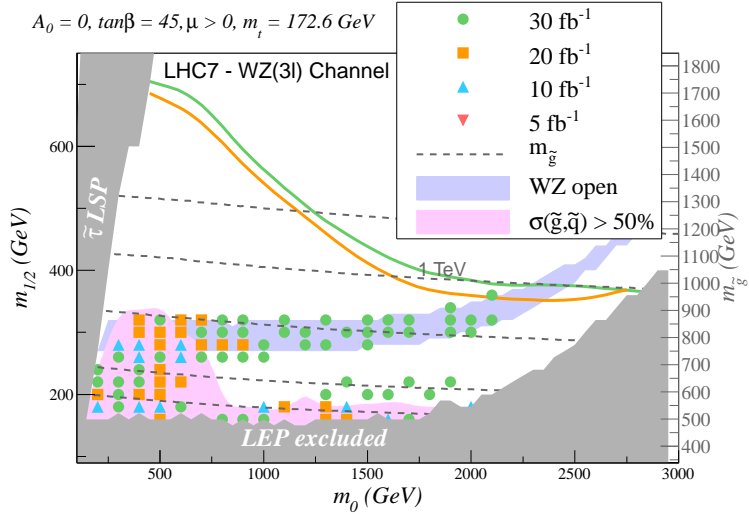,width=10cm}
\end{center}
\vspace*{-4mm}
\caption{\it 
LHC reach in the mSUGRA plane for various integrated luminosities for the $WZ + \eslt$ trilepton signal.
The pink region is where gluino and/or squark production contribute to at
least half the signal, whereas in the purple band the $WZ+\eslt$ channel
is accessible via electroweak $\tw_1\tz_2$ production. 
Below the green (orange) solid contours there will be a $5\sigma$ signal for SUSY via
the optimized jets plus $\eslt$ LHC7 search discussed in Ref.\cite{Baer:2011aa} for
30~fb$^{-1}$ (20~fb$^{-1}$).  
}\label{fig:reach-3l} 
\end{figure}

The results are shown in Fig.~\ref{fig:reach-3l}, again for four values
of integrated luminosities. All points shown are deemed visible for the
corresponding integrated luminosity. The gray regions show the parts of
the $m_0-m_{1/2}$ plane excluded by theoretical considerations or by
experimental constraints.  The purple band across the middle of the plot
shows the region in parameter space where $M_Z < m_{\tz_2} - m_{\tz_1} <
m_h$, while the pink area at low values of $m_0$ and $m_{1/2}$
corresponds to the region where at least 50\% of the signal comes from
gluino and/or squark production.
From Fig.~\ref{fig:reach-3l} we see
that, for heavy squarks ($m_0 > 800$~GeV), the signal mostly comes from electroweakly
produced inos. 
For an integrated luminosity of 5~fb$^{-1}$ no points are visible. 
However, for an integrated luminosity of 10~fb$^{-1}$, the enhancement of the signal
from gluino and squark production renders a few points at low $m_0$ and low $m_{1/2}$  accessible. 
For 20~fb$^{-1}$ the reach extends up to $m_0 \sim 800$~GeV and $m_{1/2} \sim 300$~GeV.
Finally, for 30~fb$^{-1}$, all of the region where the $WZ + \eslt$
channel is open can be probed up to $m_{1/2} \sim 350$~GeV.
In the heavy squark region ($m_0 > 800$~GeV), the signal is enhanced by $\tw_2$ and $\tz_3$
production, allowing the LHC to probe gluino masses up to 900~GeV. We point out that without the enhancement
of heavy electroweak-ino production no reach is expected even for 30~fb$^{-1}$, as shown by the red curve in Fig.~\ref{fig:plane-3l}.
We note that there are also visible points at low $m_{1/2}$, below the $M_Z <
m_{\tz_2} - m_{\tz_1} < m_h$ band, where the $\tz_2 \to \tz_1 Z$ and
$\tw_1 \to \tz_1 W$ decays are closed, but $Z$s and $W$s are still
produced from heavier EW-ino decays. 
It is also worth noting that the focus point (light higgsino) region does not 
enhance the signal. This is partly due to the more compressed chargino/neutralino 
spectrum in this region leading to softer $p_T$ and $\eslt$~\cite{Baer:2011ec,Bobrovskyi:2011jj}. 
We stress that in Fig.~\ref{fig:reach-3l}
we have only considered observability via $WZ+\eslt \to \ell^+\ell^-\ell^{'}+\eslt$
and for the region below the ``$WZ$ band'' the golden trilepton signal where the OS/SF dilepton pair has a mass below $M_Z$
can be used as a discovery channel, as shown by the dashed lines in Fig.~\ref{fig:sig-3l}.

We also show in Fig.~\ref{fig:reach-3l} the optimized LHC7 reach in the jets plus $\eslt$ channel from
Ref.~\cite{Baer:2011aa} (solid lines) for $\mathcal{L} = 20$~fb$^{-1}$
and 30~fb$^{-1}$.  These curves correspond to an optimization over
several $\eslt$ plus jets channels with zero leptons and do not include
the dedicated cuts for the $WZ+\!\eslt$ signal discussed here.  As we
can see, for such large integrated luminosities the
$WZ+\!\eslt$ trilepton channel 
is competitive with general purpose searches if squarks are essentially 
decoupled ($m_{\tilde{q}} \gtrsim 2$~TeV) and the neutralino masses
lie in the $M_Z < m_{\tz_2} - m_{\tz_1} < m_h$ band. 

%

\section{The $\ell^+\ell^- jj+\eslt$ channel}
\label{sec:lljj}

As seen in the last section, the main challenge of the trilepton signal
is its small rate, which requires relatively high luminosities for
observability. A way to increase the rates from $WZ+\eslt$ events is to
consider the dilepton channel, where $W \to jj$.  However, while the
main SM background for the $\ell^{+}\ell^{-}\ell^{'}+\eslt$ channel was
weakly produced ($WZ$), the $\ell^+\ell^- jj+\eslt$ channel has an
irreducible $t\bar{t}$ background, which can easily overcome the $WZ+\eslt$ signal 
due to its large cross section.  Nonetheless,
we will show that once evidence of a $\tw_1 \tz_2$ signal has been seen
in the trilepton channel, a corroborative signal (with lower
significance) is expected in the dilepton channel.

Using the same signal and BG event samples discussed in Sec.~\ref{sec:3l}, we extract events with a 
$\ell^+\ell^-jj+\eslt$ topology requiring: 
\begin{itemize}
\item[] \underline{Pre-Selection Cuts:}
\item $p_T(\ell )>20$ GeV and $|\eta (\ell )|<2.5$ on isolated leptons,
\item $n(jets)\ge 2$,
\item $n(b-jets)=0$,
\item $n({\rm isol.\ leptons})=2$ (of OS/SF).
\end{itemize}

In Table~\ref{tab:xsecs2}, we show the cross sections after the pre-selection cuts above
for the leading BG processes and the $\tw_1\tz_2$ signal for the same 
chargino and neutralino masses used in Table~\ref{tab:xsecs}.
As seen in the Table, after the pre-selection cuts, the SM BG is dominated by $Z+jets$, followed by $t\bar{t}$. 
To remove much of the background from $Z+jets$ production, we further require:
\begin{itemize}
\item $\eslt >40$ GeV,
\item $\eslt /M_{eff} >0.1$,
\item $\Delta\phi (\vec{p}_{jet},\vec{\eslt})>0.4$ for the three hardest
  $p_T$ jets.
\end{itemize}
After these cuts have been applied, the SM background becomes
dominated by $t\bar{t}$, which still surpass the signal by almost two
orders of magnitude, as shown in Table~\ref{tab:xsecs2}.  However, we
have not yet made use of the fact that, for the signal, the dijet
invariant mass distribution should reconstruct to $m(jj)\sim
M_W$. Therefore, in addition to the previous cuts, we include: \bi
\item $|m(jj) - M_W| < 20$~GeV,
\ei
where $m(jj)$ is the invariant mass of the two highest $p_T$ jets.

\begin{table}
\begin{center}
\begin{tabular}{|l|r|r|r|r|}
\hline
& $Z + jets$ & $t\bar{t}$ &  Total BG & Signal\\
\hline
Events Generated & 6.9M & 5.1M &  & 200K \\
Total $\sigma$ & $7.6 \times 10^6$ & $5 \times 10^3$ & $7.8 \times 10^{6}$ & $1.1 \times 10^3$ \\
\hline
Pre-selection   & 11,542 & 465 &  12,155 & 9.0 \\
$\eslt > 40$~GeV   & 71.1 & 357  & 453 & 6.4 \\
$\eslt/M_{eff} > 0.1$   & 45.8 & 345   & 415 & 6.2 \\
$\Delta\phi(j,\eslt) > 0.4$   & 31.0 & 296  & 346 & 5.3 \\
$m(jj)$ cut & 5.4 & 40.4  & 48.6 & 1.7 \\
$m(\ell^{+}\ell^{-})$ cut & 0 & 5.9 & 6.5 & 1.6 \\
\hline
\end{tabular}
\caption{Number of events generated, total cross section and cross section after cuts for the dominant backgrounds
and for the signal in the dilepton OS/SF channel.
All cross sections are in fb and the signal corresponds to $\tw_1\tz_2$ production with $m_{\tw_1} = 189.3$~GeV, $m_{\tz_2} = 187.3$~GeV and $m_{\tz_1} = 89.4$~GeV. 
The total BG values include all processes listed in the text, including the subdominant ones not shown in the Table.
\label{tab:xsecs2}}
\end{center}
\end{table}

\begin{figure}[!t]
\begin{center}
\epsfig{file=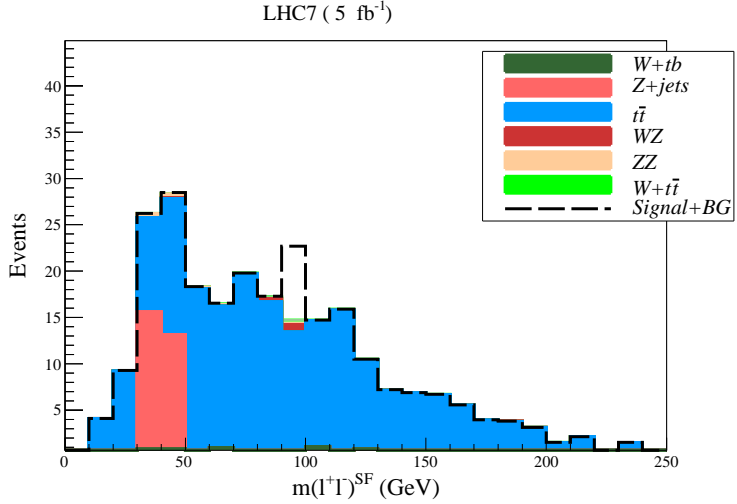,width=10cm}
\end{center}
\vspace*{-4mm}
\caption{\it 
Number of OS/SF dilepton events expected in 5 fb$^{-1}$ of LHC7 data versus $m(\ell^+\ell^- )$ for various 
summed SM backgrounds (shaded) and signal plus BG (dashed) with
 $m_{\tw_1} = 189.3$~GeV, $m_{\tz_2} = 187.3$~GeV and $m_{\tz_1} = 89.4$~GeV. 
}\label{fig:mll} 
\end{figure}

In Fig.~\ref{fig:mll}, we show the $m(\ell^+\ell^- )$ distribution for
signal and background after all the above cuts have been applied.  The
dominant backgrounds displayed are $t\bar{t}$ and $Z+jets$ (including
$Z\to\tau\bar{\tau}$).  Due to the $\eslt$ cut, the remaining $Z+jets$
contribution comes mostly from $Z\to\tau\bar{\tau}$, with $\tau$s
decaying leptonically. Therefore all $Z+jets$ events have
$m(\ell^+\ell^-) < M_Z$.  For these dominant backgrounds-- $t\bar{t}$,
$Z\to\tau\bar{\tau}$ etc.-- we expect nearly equal contributions of
opposite-flavor dileptons (OF): $e^\pm\mu^\mp$ pairs, while signal is
all in the SF dilepton channel.  Hence the OF distribution can serve as a
background normalization.  As seen in Fig.~\ref{fig:mll}, the signal is
visible over the $t\bar{t}$ distribution at $m(\ell^+ \ell^-) =
M_Z$.\footnote{Isajet does not include the $Z$ width smearing in the
real $Z$ emission, so all $Z\to \ell^+\ell^-$ fall exactly at $M_Z$.}
Therefore, after applying the cuts listed
above, we also require: \bi
\item $|m(\ell^{+}\ell^{-}) - M_Z| < 10$~GeV .
\ei

%

%
%

\begin{figure}[!t]
\begin{center}
\epsfig{file=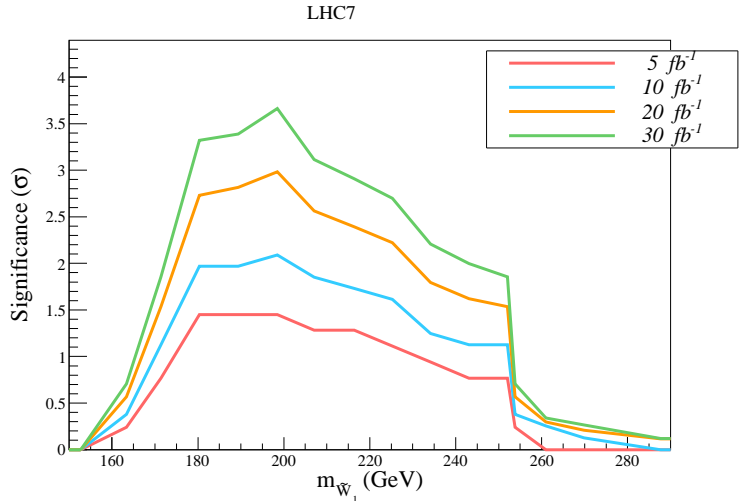,width=10cm}
\end{center}
\vspace*{-4mm}
\caption{\it 
Significance of $\tw_1\tz_2\to WZ+\!\eslt\to jj+(Z\to\ell^+\ell^-)+\!\eslt$ signal 
for various integrated luminosities at LHC7, after the cuts listed in Table~\ref{tab:xsecs2} have been applied. 
}\label{fig:sig-2l} 
\end{figure}

As shown in Table~\ref{tab:xsecs2}, after the $m(\ell^+\ell^- )$ cut has
been included, the BG is almost entirely given by $t\bar{t}$, which
contribution can be estimated using the opposite-flavor dilepton
invariant mass, as mentioned above. However, although the dilepton
signal rate is considerably superior to the trilepton case, the signal
still is significantly below the background.  In Fig.~\ref{fig:sig-2l}
we plot the signal significance after the above cuts have been applied
for various integrated luminosity values versus $m_{\tw_1}$.  As in
Fig.~\ref{fig:sig-3l} we assume a mSUGRA line with $m_0=10$ TeV,
$A_0=-2m_0$, $\tan\beta =25$ and $\mu >0$.  We see immediately that the
significance in the dilepton channel is almost half of the significance
in the trilepton channel, shown in Fig.~\ref{fig:sig-3l}.  Nevertheless,
corroborative evidence at the $2\sigma$ level is expected over almost
the entire kinematically allowed range for an integrated luminosity of
20-30~fb$^{-1}$.

\section{Summary and conclusions}
\label{sec:conclude}

In this paper, we have pointed out that for a class of SUSY models with
decoupled matter scalars, $m_{\tw_1} \sim m_{\tz_2} \lesssim m_{\tg}/3$
and gluino masses above $\sim$500 GeV, electroweak production of
$\tw_1\tz_2$ dominates the SUSY production rate at LHC7.  
We have examined the case where $M_Z< m_{\tz_2}-m_{\tz_1}<m_h$, 
for which we expect the two-body decay modes
$\tz_2 \to \tz_1 Z$ and $\tw_1\to\tz_1 W$ to dominate, leading to rather
simple final state topologies including $(Z\to\ell^+\ell^-)+(W\to
\ell'\nu_{\ell'})+\eslt$ (trileptons) and $(Z\to\ell^+\ell^-)+(W\to
q\bar{q}')+\eslt$ (dilepton plus jets).

Evaluation of the trilepton signal against SM backgrounds shows that the
SUSY signal should be observable with a $5\sigma$ significance at LHC7
up to $m_{\tw_1}\sim 250$ GeV (depending on $m_{\tz_1}$), for an integrated
luminosity of 30~fb$^{-1}$.  In models with gaugino
mass unification, this corresponds to a range in gluino masses of
$m_{\tg}\sim 700-900$ GeV. 
Moreover, we find that for most of this region a $\sim 2\sigma$
excess is expected in the 5 fb$^{-1}$ data sample that has already been
accumulated.
Assuming 30~fb$^{-1}$ of integrated luminosity at LHC7, the
trilepton channel will be competitive in reach with the canonical
multijet plus $\eslt$ search in models with unified gaugino mass
parameters. 
If a signal is seen in the trilepton channel, a $2-3.5\sigma$ confirmatory signal is also 
expected in the dilepton plus jets channel for most of the parameter
space, thus making a stronger case for the $\tw_1 \tz_2$ signal.
Most importantly, the simultaneous presence of these
signals will point to the SUSY origin of any new physics that might be
discovered in the 2012 run.

\section*{Acknowledgments} 

This work was supported in part by the U.S. Department of Energy under grant 
Nos.~DE-FG02-04ER41305, DE-FG02-04ER41291 and DE-FG02-95ER40896, 
by the IN2P3 of France under contract PICS FR-USA No.~5872 
and by the Fundac\~ao de Apoio \`a Pesquisa do Estado de S\~ao Paulo (FAPESP).



\begin{thebibliography}{99}
\small
%
\bibitem{wss} 
For a review of SUSY, see
H.~Baer and X.~Tata, {\it Weak Scale Supersymmetry: From 
Superfields to Scattering Events}, 
(Cambridge University Press, 2006).
\bibitem{atlas} 
ATLAS collaboration, G. Aad {\it et al.}, 
arXiv:1109.6572, arXiv:1109.6606, JHEP 11 (2011) 99, and arXiv:1110.6189; 
see also\\ {\tt https://twiki.cern.ch/twiki/bin/view/AtlasPublic/SupersymmetryPublicResults}
%
\bibitem{cms} 
CMS collaboration, S. Chatrchyan {\it et al.}, \prl{107}{2011}{221804}, CMS-PAS-SUS-11-008; 
see also {\tt https://twiki.cern.ch/twiki/bin/view/CMSPublic/PhysicsResultsSUS}
%
\bibitem{msugra} 
A.~Chamseddine, R.~Arnowitt and P.~Nath, 
Phys. Rev. Lett. {\bf 49}, 970 (1982);
R.~Barbieri, S.~Ferrara and C.~Savoy, 
Phys. Lett. B{\bf 119}, 343 (1982); N.~Ohta, Prog. Theor. Phys. {\bf
  70}, 542 (1983);
L.~J.~Hall, J.~Lykken and S.~Weinberg, Phys. Rev. {\bf D27}, 2359 (1983);
For a review, see {\it e.g.} R. Arnowitt and P. Nath, arXiv:0912:2273 (2009).
%
\bibitem{atlasSO10} 
ATLAS collaboration, ATLAS-CONF-2011-098.
%
\bibitem{bcpt1} H.~Baer, C.~H.~Chen, F.~Paige and X.~Tata, \prD{52}{1995}{2746}.
%
\bibitem{bcpt2} H.~Baer, C.~H.~Chen, F.~Paige and X.~Tata, \prD{53}{1996}{6241}.
%
\bibitem{cascade} H. Baer, V. Barger, D. Karatas and X. Tata, \prD{36}{1987}{96}; 
H.~Baer, R.~M.~Barnett, M.~Drees, J.~F.~Gunion, H.~E.~Haber, D.~L.~Karatas and X.~R.~Tata,
Int.\ J.\ Mod.\ Phys.\  A {\bf 2}, 1131 (1987);
H. Baer, A. Bartl, D. Karatas, W. Majerotto and X. Tata, \ijmpa{4}{1989}{4111};
H.~Baer, X.~Tata and J.~Woodside, \prD{42}{1990}{1568};
for earlier work on sparticle decays to just gauginos, see 
H. Baer, J. Ellis, G. Gelmini, D. V. Nanopoulos and X. Tata, \plb{161}{1985}{175};
G. Gamberini, \zpc{30}{1986}{605}; H. Baer and E. Berger, \prd{34}{1986}{1361}.
%
\bibitem{btw} H. Baer, X. Tata and J. Woodside, \prd{41}{1990}{906}
and \prd{45}{1992}{142}.
%
\bibitem{dine} M. Dine, A. Kagan and S. Samuel, \plb{243}{1990}{250};
  N.~Arkani-Hamed and H.~Murayama, \prd{56}{1997}{6733}.
%
\bibitem{ckn} M.~Drees, \prd{33}{1986}{1468}; S.~Dimopoulos and
  G.~Giudice, \plb{357}{1995}{573}; A.~Pomarol and D.~Tomassini,
  \npb{466}{1996}{588};  A. Cohen, D. B. Kaplan and A. Nelson,
  \plb{388}{1996}{588}; 
%
\bibitem{Sekmen:2011cz} 
S.~Sekmen, {\it et al.},  arXiv:1109.5119 [hep-ph].

%
\bibitem{prospino} W. Beenakker, R. Hopker, M. Spira, hep-ph/9611232 (1996).
%

%
\bibitem{Mrenna:2011ek} 
S.~Mrenna [For the CMS Collaboration],  arXiv:1110.4078 [hep-ph].
%
\bibitem{wh} H. Baer, V. Barger, A. Lessa, W. Sreethawong and X. Tata,
arXiv:1201.2949.
%
\bibitem{trilepton} H. Baer, C. H. Chen, F. Paige and X. Tata,
\prd{50}{1994}{4508}.
%
\bibitem{supp} H.~Baer and X.~Tata, \prd{47}{1993}{2739}. 

%
\bibitem{isajet} F. Paige, S. Protopopescu, H. Baer and X. Tata,
hep-ph/0312045.
%
\bibitem{alpgen} M. Mangano, M. Moretti, F. Piccinini, R. Pittau and
A. Polosa, \jhep{0307}{2003}{001}.
%
%
\bibitem{madgraph} J. Alwall, M. Herquet, F. Maltoni, O. Mattelaer and
T. Stelzer, \jhep{1106}{2011}{128}.
%
%
\bibitem{pythia} T. Sjostrand, S. Mrenna and P. Skands,
\jhep{0605}{2006}{026}.

%
\bibitem{mcfm}
MCFM, by J. Campbell and R. K. Ellis. See R.~K.~Ellis,
  Nucl.\ Phys.\ Proc.\ Suppl.\  {\bf 160} (2006) 170.
%
\bibitem{bbm} H. Baer, V. Barger and A. Mustafayev, arXiv:1112.3017.
%
\bibitem{Baer:2011aa}
  H.~Baer, V.~Barger, A.~Lessa and X.~Tata,
  arXiv:1112.3044 [hep-ph].

\bibitem{Baer:2011ec} 
  H.~Baer, V.~Barger and P.~Huang,
   \jhep{111}{2011}{031}.
  
\bibitem{Bobrovskyi:2011jj} 
  S.~Bobrovskyi, F.~Brummer, W.~Buchmuller and J.~Hajer,
  arXiv:1111.6005 [hep-ph].
    
\end{thebibliography}
\end{document}